\documentclass[twocolumn,pre,aps,10pt,showpacs,superscriptaddress]{revtex4-1}

\usepackage{graphicx}
\usepackage{amssymb,amsfonts,amsmath}
\usepackage{epstopdf}

\usepackage{color} 

    \usepackage{multirow}
    
    \newcommand{\nesg}{{\sc nesg} }
    
\begin{document}

\title{Statistical analysis of crystallization database links protein physicochemical features with crystallization mechanisms.}

\author{Diana Fusco}
\affiliation{Program in Computational Biology and Bioinformatics, Duke University, Durham, NC 27708}
\affiliation{Department of Chemistry, Duke University, Durham, NC 27708}

\author{Timothy J. Barnum}
\affiliation{Department of Chemistry, Duke University, Durham, NC 27708}
\affiliation{Department of Chemistry, Massachusetts Institute of Technology, Cambridge, MA, 0213}

\author{Andrew E. Bruno}
\affiliation{Center for Computational Research, State University of New York, Buffalo, NY 14260}

\author{Joseph R. Luft}
\affiliation{Hauptman-Woodward Medical Research Institute, Buffalo, NY 14203}

\author{Edward H. Snell}
\affiliation{Hauptman-Woodward Medical Research Institute, Buffalo, NY 14203}
\affiliation{Department of Structural Biology, State University of New York, Buffalo, NY 14203}

\author{Sayan Mukherjee}
\affiliation{Department of Statistical Science, Duke University, Durham, NC 27708}
\affiliation{Department of Computer Science, Duke University, Durham, NC 27708}
\affiliation{Department of Mathematics, Duke University, Durham, NC 27708}

\author{Patrick Charbonneau}
\affiliation{Department of Chemistry, Duke University, Durham, NC 27708}
\affiliation{Department of Physics, Duke University, Durham, NC 27708}

\begin{abstract}
X-ray crystallography is the predominant method for obtaining atomic-scale information about biological macromolecules. Despite the success of the technique, obtaining well diffracting crystals still critically limits going from protein to structure. In practice, the crystallization process proceeds through knowledge-informed empiricism. Better physico-chemical understanding remains elusive because of the large number of variables involved, hence little guidance is available to systematically identify solution conditions that promote crystallization. 
To help determine relationships between macromolecular properties and their crystallization propensity, we have trained statistical models on  samples for 182 proteins supplied by the Northeast Structural Genomics consortium. Gaussian processes, which capture trends beyond the reach of linear statistical models, distinguish between two main physico-chemical mechanisms driving crystallization. One is characterized by low levels of side chain entropy and has been extensively reported in the literature. The other identifies specific electrostatic interactions not previously described in the crystallization context. Because evidence for two distinct mechanisms can be gleaned both from crystal contacts and from solution conditions leading to successful crystallization, the model offers future avenues for optimizing crystallization screens based on partial structural information. The availability of crystallization data coupled with structural outcomes analyzed through state-of-the-art statistical models may thus guide macromolecular crystallization toward a more rational basis.

\end{abstract}
 
\maketitle

\section{Introduction}

X-ray crystallography is the most frequently used technique to obtain structural information about biological macromolecules, currently accounting for more than 88\% of the entries in the Protein Data Bank (PDB)~\cite{PDB}. However, as its name suggests, the method fundamentally relies on obtaining well diffracting crystals of the macromolecules or complexes of interest (generally termed proteins in the crystallographic context and used as such throughout this paper). Some quantitative data on the success rate of crystallization comes from the Protein Structure Initiative (PSI). This program, initiated by the National Institutes of Health, has enabled high-throughput structural studies of biomolecules that track the experimental outcome, success or failure. Analysis of this data reveals that despite the large scale efforts, fewer than 30\% of the proteins that are expressed and purified yield diffraction quality crystals, and of these only 67\% provide structures (20\% of the expressed and purified samples)~\cite{chen:2004,terwilliger:2009}. There are currently about 100,000 structures in the PDB~\cite{PDB}, but more than 10 million non-redundant protein chain sequences have been reported~\cite{pruitt:2005}. The large number of proteins for which detailed structural knowledge remains unavailable is an ongoing challenge for high-throughput crystallization.

The current approach to crystallization is empirical. Proteins are screened against arrays of many chemical conditions that are biologically ``friendly" and have yielded crystals in the past~\cite{mcpherson:1999}. As an example, the High-Throughput Crystallization Screening (HTS) laboratory at the Hauptman Woodward Medical Research Institute uses 1,536 different chemical conditions each aiming to reduce protein solubility so as to obtain ordered crystallization~\cite{luft:2011}. The approach yields at least one crystal in about 50\% of the samples~\cite{snell:2008}, but, from tracking the PSI supplied samples, only about half of those initial crystal hits go on to yield subsequent structural information. This success rate is respectable in the field, although it should be noted that the result is not a per cocktail (crystallization experiment) statistic, but a binary analysis on the presence of a crystal within 1,536 different experiments. Out of all the screening performed in the HTS laboratory it is estimated that only 0.2\% of individual crystallization screening conditions yield a crystal; failure is unfortunately all too common~\cite{newman:2012}. One may thus hope that an improved physico-chemical understanding of protein crystallization could help navigate the chemical screening space more efficiently~\cite{rupp:2004}.


The positive and negative outcome data captured by the PSI and similar structural genomics (SG) efforts have been employed to determine the key factors that affect protein crystallization.  
 A number of studies have used amino acid sequence features as inputs to machine-learning classification schemes, in order to identify proteins that should easily crystallize and thus be good SG targets~\cite{smialowski:2006,slabinski:2007b,kurgan:2009,zucker:2010,mizianty:2011}. This strategy faces two main difficulties. First, it typically relies on a protein's amino acid sequence, which is only indirectly related to crystal assembly. Surface residues are more directly linked to the crystallization process~\cite{dale:2003}, but are challenging to determine with high fidelity~\textit{de novo}. Second, typical machine-learning methods based on Support Vector Machine, which divide the feature space between different classes of macromolecules~\cite{cristianini:2000}, are deterministic and can be hard to interpret physically. The complexity of the function that separates positive and negative regions of parameter space  typically hinders the physico-chemical interpretation of the results and thus the transfer of microscopic insights to applications beyond crystal formation, such as peptide design~\cite{rupp:2004,kurgan:2009,Saven2010,boyle:2011}. 
 
 Two statistical inference models trained on a richer set of protein features have gone beyond these difficulties~\cite{price:2009,derewenda:2009}. Both of them find that low values of surface side chain entropy (related to the degrees of freedom of the surface residues) and a high fraction of small surface residues, such as glycine and alanine, assist crystallization. They thus support surface entropy reduction (SER) mutagenesis, which broadly prescribes replacing large residues, e.g. lysines and glutamic acids, with alanines~\cite{derewenda:2004,derewenda:2010}, in order to facilitate crystallization. Yet the two modeling approaches use somewhat orthogonal algorithms for predicting crystallization propensity and their results do not always agree~\cite{kurgan:2009}.

Part of this discrepancy may come from the linearity of the underlying models. Linear models have the advantage of being easily interpretable, but they struggle to capture subtle non-linear and possibly non-monotonic trends, which can make them sensitive to the details of the training set. However, protein crystallization responds non-linearly to changes in solution conditions. Extremely low values of side chain entropy indeed hinder crystallization by compromising protein solubility, as observed in experiments ~\cite{cooper:2007} and as predicted in solubility models~\cite{Price:2011}. George and Wilson also carefully documented the non-linearity of protein crystallization by identifying the range of second virial coefficient (not too high, not too low) over which proteins typically crystallize~\cite{george:1994}, a result that is fundamental to the materials physics understanding of protein assembly~\cite{rosenbaum:1996,Wolde1997,Bianchi2011,fusco:2013a,fusco:2013b}. 

In this work, we have used a subset of the screening results from the North East Structural Genomics consortium (\nesg) to train models based on Gaussian processes (GP). GP replace specific constraints on the functional form of the model with a prior distributions that weighs all of the (infinite) smooth functions~\cite{GP}, and can thus better capture the non-linear and non-monotonic relations in a dataset. 
The resulting models help us address two fundamental questions about protein crystallization. (i) What protein properties determine crystallization propensity in standard screens? (ii) How do these properties relate to successful crystallization conditions? Answering (i) enhances existing mutagenesis prescriptions to facilitate the crystallization of recalcitrant proteins without denaturing their structure; answering (ii) suggests guidelines for tailoring and narrowing the set of solution conditions for crystallizing a given protein.

\section{Results and Discussion}
\begin{figure*}[htb]
\centering
\includegraphics[width=0.9\textwidth]{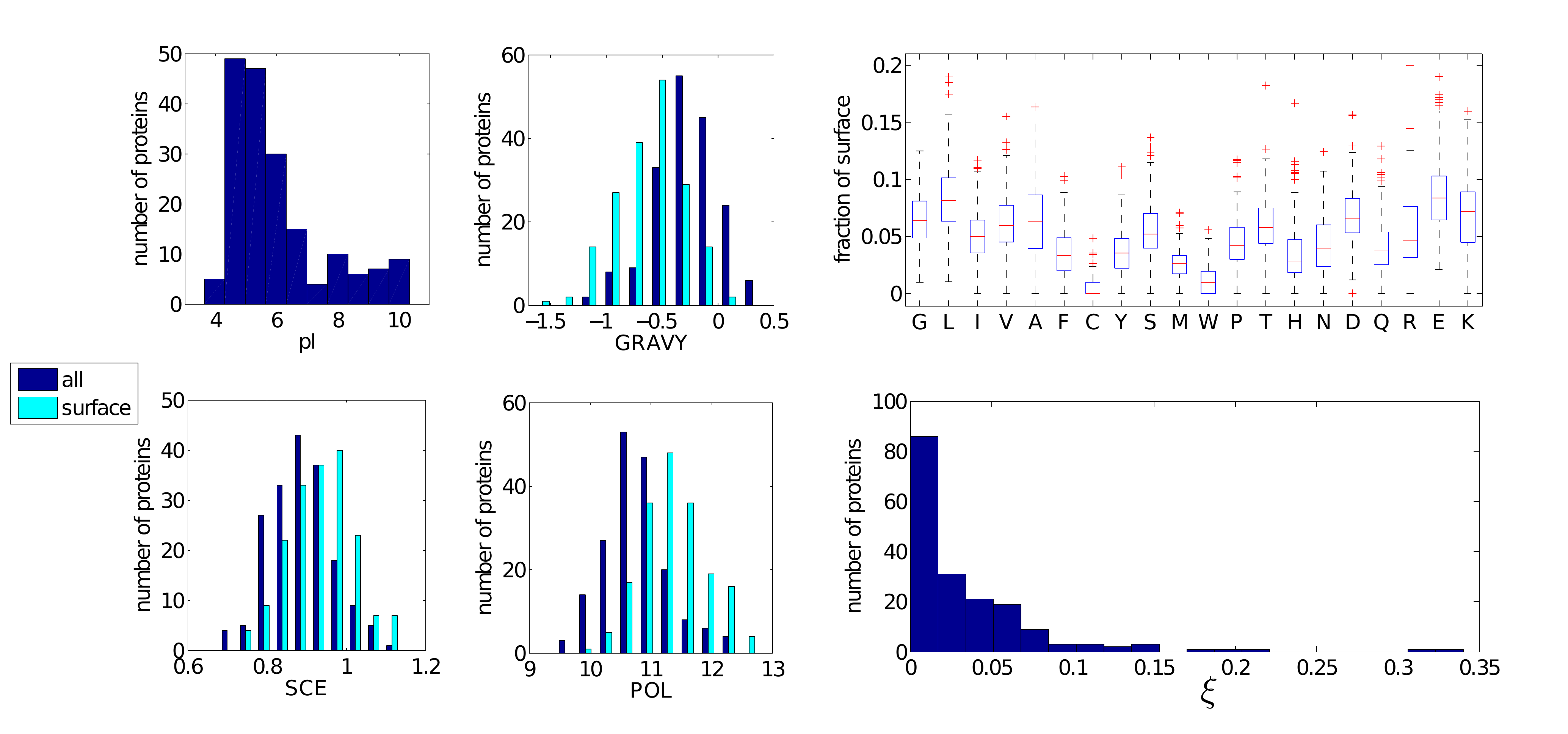}
\caption{Summary of some of the features of the proteins contained in the \nesg\ dataset. Histograms report the pI, average hydropathy index (GRAVY), average side chain entropy (SCE), average polarity (POL). Both the overall (blue) and the surface (cyan) valus are shown for the GRAVY index, side chain entropy (SCE) and polarity (POL). Distribution of the surface coverage for each amino acid and histogram of the crystallization propensity $\xi$ are also shown.}
\label{fig:protein}
\end{figure*}

The dataset provided by the \nesg\ contains information about 182 distinct proteins that were supplied in a common buffer and set up against an array of 1,536 different chemical cocktails representing an extensive set of crystallization conditions (see Methods). The different microbatch under-oil experiments were imaged over time and each outcome was visually classified as containing a crystal or not. Protein structures were subsequently determined using X-ray crystallography by the \nesg.
In this dataset, a broad range of crystallization propensity, $\xi$, defined as the fraction of the 1,536 cocktails that successfully generated crystals, is observed. Two proteins formed crystals in as many as 30\% of the tested conditions, but most did so in only a few of the solutions (Fig.~\ref{fig:protein}). The binary classification between crystal or no crystal does not distinguish between the stochastic nucleation process and the crystal growth process, once nucleation has occurred. Both must have happened to produce a crystal. While we may have false negatives that could be reduced by replication of the crystallization screening process, the large range of related chemical conditions and the fairly large number of samples studied should largely mitigate this effect. Some proteins may nucleate more easily than others, but once nucleation occurs, growth follows and the dataset records this outcome. Nucleation could perhaps be deconvoluted from crystallization by recording the number of crystals produced per chemical condition, but it has not been attempted for this study. The term crystallization in our case therefore necessarily indicates both crystal nucleation and crystal growth. 

\subsection{GPR: Crystallization propensity}

\begin{figure*}[htb]
\centering
\includegraphics[width=\textwidth]{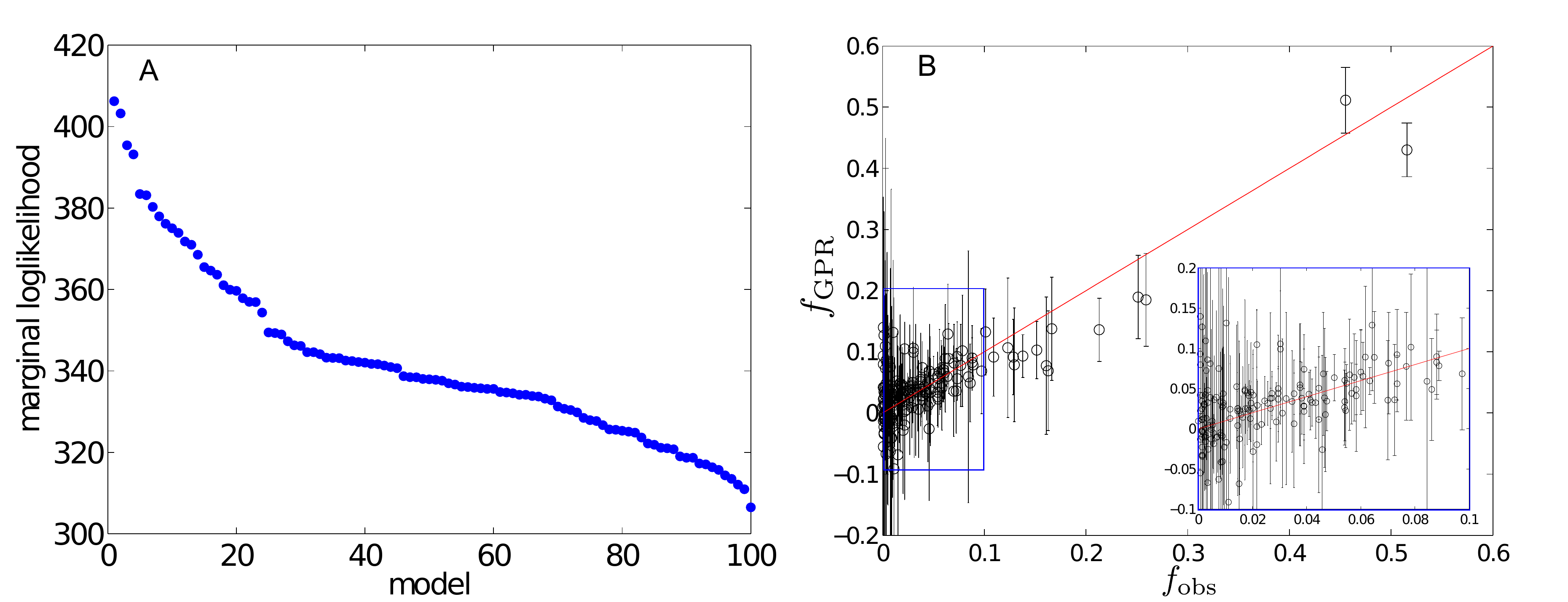}
\caption{A: Marginal log likelihood for 100 models that each are distinct local maxima of Eq.~(\ref{eq:marginal_likelihood}). The model with the highest value is used for the rest of the analysis. B: Scatter plot of the observed output function $f_{\mathrm{obs}}=\frac{\xi}{1-\xi}$ and its value predicted by the GPR model $f_{\mathrm{GPR}}$ using a LOO cross validation with 95\% confidence intervals. The inset details the low propensity data.}
\label{fig:LOOCV}
\end{figure*}

\begin{figure*}[htb]
\centering
\includegraphics[width=\textwidth]{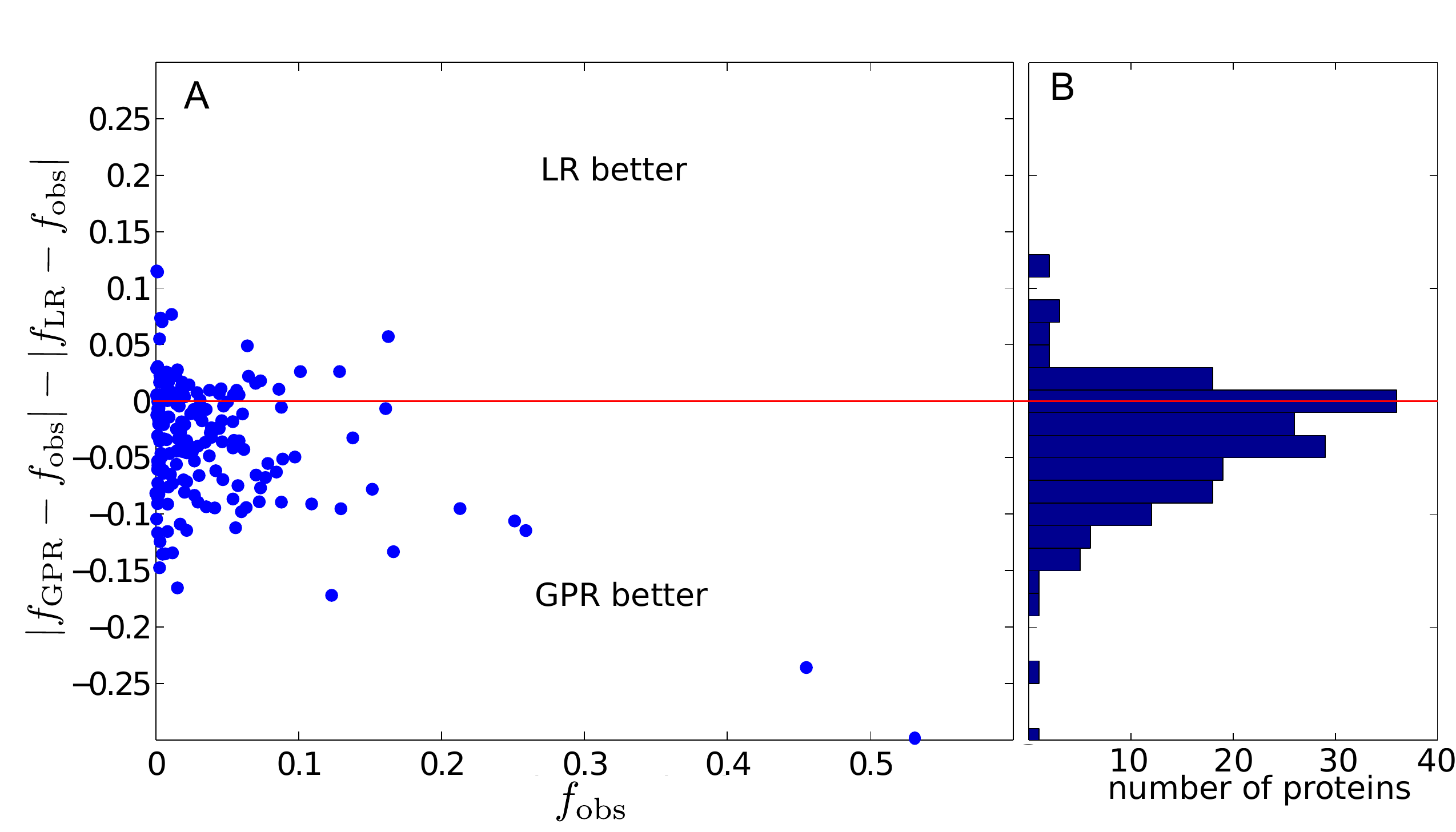}
\caption{Difference between the observed and the modeled propensity values $f$. A: Points in the upper half of the graph correspond to the cases in which LR performs better than GPR, and vice versa for the lower half. B: The histogram summarizes the overall performance of LR and GPR.}
\label{fig:GP_LR}
\end{figure*}

We identified some of the factors behind facile protein crystallization by training a Gaussian process regression (GPR) model for $\xi$ using a set of physico-chemical properties as predictive variables $\mathbf{x}$. (For mathematical convenience the output function $f(\mathbf{x})$ of the GPR model is chosen to be $f=\xi/(1-\xi)$ (instead of $f=\xi$), but the uniqueness of this transformation and of its inverse, $\xi=f/(1+f)$, results in no loss of generality. See also Methods for more details.)
The flexibility of GP enables GPR models to capture any continuous relationship, no matter how complex, between $\mathbf{x}$ and the output function $f$. In order to avoid overfitting the model parameters, we optimized the model marginal log-likelihood (Eq.~(\ref{eq:marginal_likelihood})), which rewards good fitting of the data while penalizing overly complex models. The training process selects one specific function that best captures the effect of the predictive variables $\mathbf{x}$.
Because many local maxima of the marginal log-likelihood can be found over the parameter space, a broad search is necessary. An exhaustive sampling is out of computational reach, but the largest maximum we located is also the best performer in leave-one-out (LOO) cross-validation (Fig.~\ref{fig:LOOCV} A and B). This standard diagnostic tool for overfitting~\cite{GP}  indicates that the choice of parameters is reasonably representative of the best model. A direct comparison reveals that the resulting GPR model recovers the observed data more precisely than linear regression (LR) in 74\% of the \nesg\ proteins, with GPR performing consistently better for proteins with a moderate-to-high crystallization propensity (Fig.~\ref{fig:GP_LR}). We thus confirm that a non-linear model better relates a protein's crystallization propensity to its physico-chemical properties. 

The significance of specific predictive variables in a GPR model can be assessed by the
magnitude of their corresponding learned length scale $l$ in the GP kernel function (see Methods). A small $l$ indicates that the model has a high sensitivity to a specific property, and vice versa. In that sense $l$ plays a role similar to a weight in a LR model, but is unsigned. Determining whether a variable is positively or negatively correlated with the output of the model requires a local analysis of its predictions.

\begin{figure*}[htb]
\centering
\includegraphics[width=\textwidth]{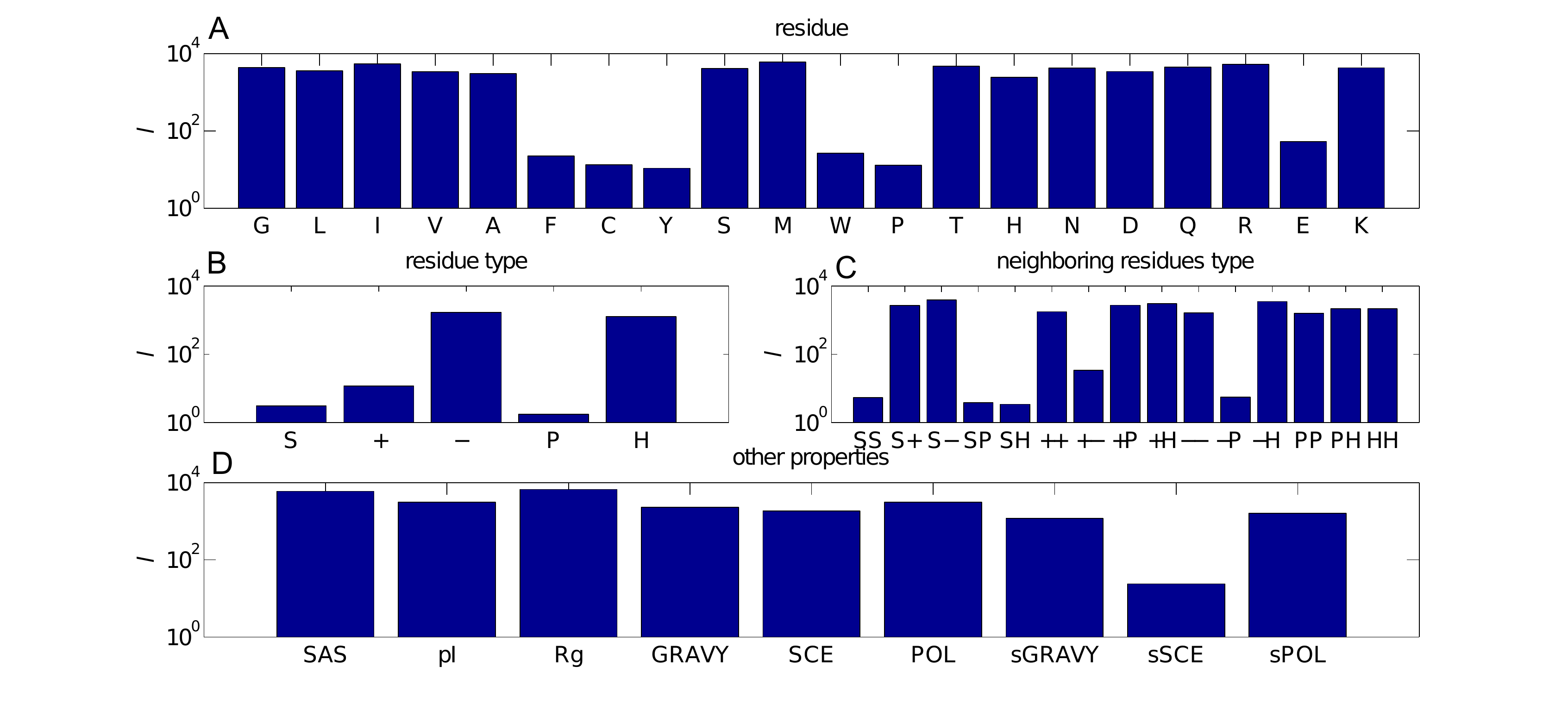}
\caption{Length scales $l$ associated with the parameters for the maximal log-likelihood model (highest point in Fig.~\ref{fig:LOOCV} A). The plots divide the explanatory variables in (A) residues, (B) residue type (small S, positively charged +, negatively charged $-$, polar P and hydrophobic H), (C) neighboring residue type (pairs of these symbols), and (D) other global protein properties (see Methods for details).}
\label{fig:length_scales}
\end{figure*}

Comparing $l$ for the different protein surface \emph{residues}  indicates that the most significant residues are the aromatics (phenylalanine (F), tyrosine (Y), tryptophan (W), and proline (P)) as well as cysteine (C) and glutamic acid (E) (Fig.~\ref{fig:length_scales}). The importance of phenylalanine and glutamic acid was uncovered in earlier studies~\cite{price:2009}, but that of the other aromatic residues and of cysteine had previously gone undetected. The contrast between the (small) $l$ associated with aromatic residues, which are hydrophobic and large, and the (large) $l$ associated with small hydrophobic residues, i.e., leucine (L), isoleucine (I), and valine (V), further indicates that only \emph{large} hydrophobic residues play a significant role, which is not all together surprising based on hydrophobicity arguments~\cite{chandler:2005}. The case of cysteine is interesting. A recent protein crystallization engineering study found that replacing some residues with cysteines promotes crystal formation because of the residue's ability to form disulfide bonds and hence dimerize~\cite{Banatao:2006}. Yet that very reactivity can also result in noncrystalline aggregation~\cite{Eiler:2001}. The non-monotonicity resulting from the two competing behaviors, which the GPR model here detects, may explain why earlier LR-based studies had not detected its importance~\cite{price:2009}.

At a coarser level, three surface residue \emph{categories} -- small, positively charged, and polar -- are found to be significant  (Fig.~\ref{fig:length_scales}). Small residues enable the formation of favorable inter-protein backbone contacts and their low side chain entropy eases the formation of crystal contacts~\cite{price:2009, derewenda:2009}. The general importance of surface side chain entropy (sSCE) further supports this interpretation (see Method for sSCE definition). The role of the other two residue categories is more controversial. Cie\'slik and Derewenda found that polarity strongly affects whether a residue belongs to a crystal contact~\cite{derewenda:2009}, but Price \textit{et al.} did not detect any significant contribution from individual polar residues~\cite{price:2009}. Charged residues also have an ambiguous role. Lysine is thought to inhibit crystallization~\cite{price:2009}, while arginine has been suggested to facilitate crystallization in isolated instances~\cite{Dasgupta:1997,derewenda:2006,fusco:2013b}. Yet no correlation between arginine and crystallization propensity had thus far been noted.
 
The asymmetry between positively and negatively charged residues in affecting crystallization propensity is puzzling, especially because positive and negative residues are almost identically distributed over the surface of the proteins studied. One possible explanation is that the effect mirrors the asymmetry  (slightly) favoring pH$<7$ in the 1,536 condition screens (see Methods). 
This imbalance may neutralize the net charge of negative residues and therefore their ability to electrostatically affect inter-protein interactions. Another possible explanation comes from the asymmetry in water's charge distribution, strengthening interactions between water molecules and negative residues, and therefore favoring residue solvation~\cite{dill:2005}. The increased participation of negatively-charged glutamic and aspartic acids compared to positively-charged residues in protein-protein interactions bridged by water supports this second scenario~\cite{rodier:2005}, but overall the evidence remains inconclusive.

The role of \emph{neighboring pairs} of surface residues, although presumed to be significant~\cite{crystalp2}, had not been previously directly assessed. We find that small-small (SS), small-polar (SP), and small-hydrophobic (SH) pairs as well as negative-positive (+$-$) and negative-polar ($-$P) pairs markedly affect crystallization propensity (Fig.~\ref{fig:length_scales}). The significance of the first three pair types reflects the enhanced role of small residues when coupled with specific residue types in promoting backbone-backbone (small-small), side chain-backbone (small-polar) and hydrophobic (small-hydrophobic) interactions. The importance of neighboring negative and positive residues is particularly interesting. Intra-chain pairing of these two residues can indeed suppress a potential source of favorable electrostatic inter-chain interactions. This mechanism has even been observed to hinder crystal formation in computational studies~\cite{fusco:2013b}, pointing to a potential target for mutagenesis. 

Similarly to the case of individual charged residues, which shows an asymmetry between positive and negative side chains, negative-polar neighboring residues appear important, whereas positive-polar pairs are found to play no significant role. Proteins with a high crystallization propensity are significantly depleted in both of these pairings (see below and Table~\ref{tab:easy}), suggesting that both hinder crystallization. We note, however, that retraining the GPR model without the positive residue category gives as much importance to positive-polar pairs as to negative-polar pairs. Similarly, retraining the model without the negative-polar category gives added importance to the negative residue category. These observations imply that a strong correlation exists between positive residues and positive-polar pairs as well as between negative residues and negative-polar pairs. 
Yet the correlation is incomplete. The importance of positive residues is indeed better captured when they are considered alone ($10^6$ times more likely than the alternate model), and that of negative residues is better captured when they are coupled with polar residues ($10^2$ times more likely than the alternate model). The source of this correlation and asymmetry  remains unclear, and to the best of our knowledge, no physico-chemical explanation for this observation has thus far been suggested. 

None of the other protein properties, including their isoelectric point (pI), significantly affect crystallization, which is in line with a recent LR analysis of a similar dataset~\cite{price:2009}. Although  this finding may seem surprising based on earlier reports that found the pI to be an important physical factor in protein crystallization~\cite{mcpherson:1999,kantardjieff:2004,slabinski:2007}, the discrepancy likely results from the selection bias of standard screens (like those analyzed here) in favor of conditions that are expected to reduce protein solubility for most proteins.
These screens avoid pairing low-salt and extreme-pH conditions, which result in unscreened similarly charged proteins. Were these conditions present, they would likely statistically emphasize the physical importance of pI in protein crystallization. 

It is important to note that the GPR model is able to capture \textit{all} the significant trends spotted by previous LR models~\cite{price:2009}. In particular, it identifies the role of alanine and glycine (as small residues), of phenylalanine, and of sSCE in promoting crystallization. The other variables that were identified as important in the LR model of Ref.~\cite{price:2009} but are not singled out by the GPR model, such as lysine and sGRAVY,  were actually found to be redundant because of their strong correlation to sSCE~\cite{price:2009}. This result highlights the elegance with which GPR handles correlations among the explanatory variables.

\begin{figure*}[htb]
\centering
\includegraphics[width=\textwidth]{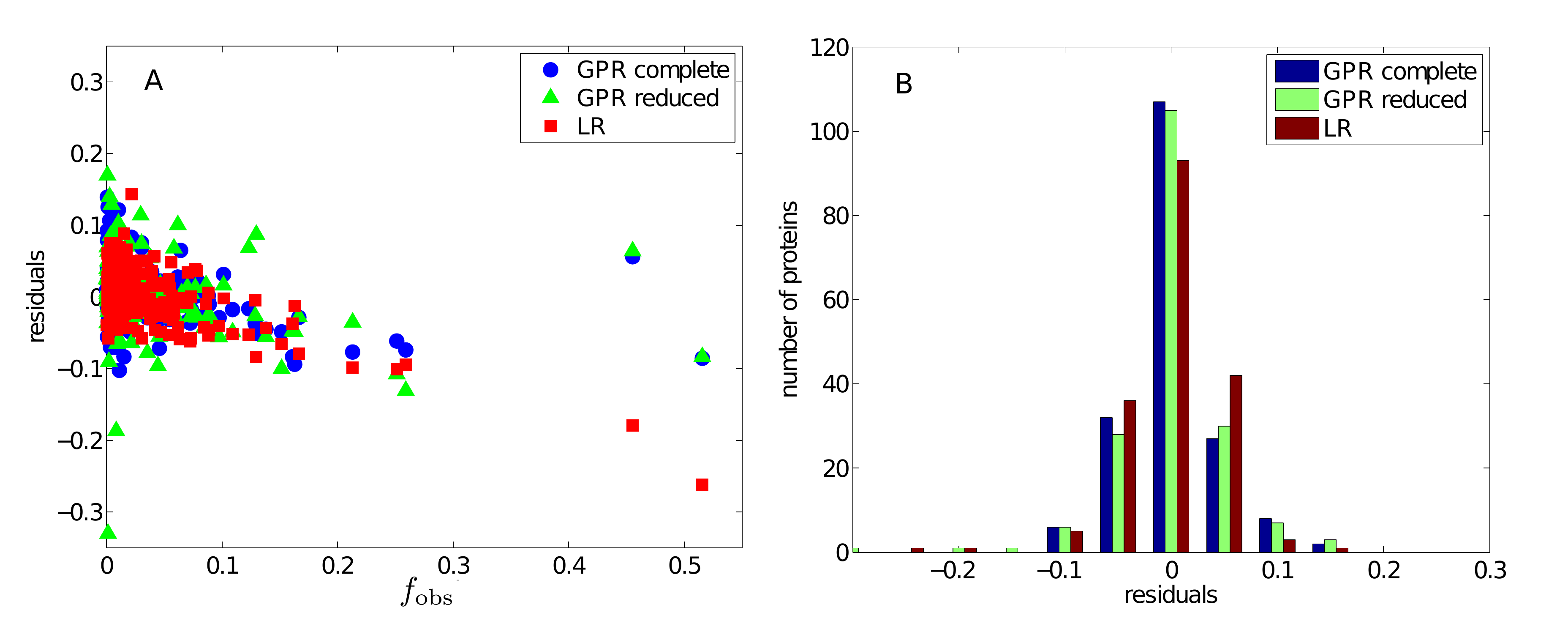}
\caption{A: Scatter plot of the regression residuals as a function of the observed output function $f_{\mathrm{obs}}$ for the complete GPR model (blue circles), the reduced GPR model with only residue categories as variables (green triangles), and the standard LR model (red squares). B: Histogram of the residuals in A that quantifies the quality of the GPR fits.}
\label{fig:residuals}
\end{figure*}

In this respect, one correlation that is inherent to our choice of variables is that between the identity of specific residues and the residue category to which they belong. Because we find that residue categories impact the crystallization propensity more significantly than most individual residues, we trained a second GPR model using only surface residue categories and pairs of surface residue categories as descriptive variables. This reduced GPR model performs very similarly to the complete version, and is also much better than LR (in 72\% of the cases, Fig.~\ref{fig:residuals}).
This analysis suggests that a coarsened description employing only residue categories could serve as a first approximation to understanding and tuning protein crystallization using mutagenesis.

\subsection{GPR: Independent crystallization mechanisms}

\begin{figure}[htb]
\centering
\includegraphics[width=0.5\textwidth]{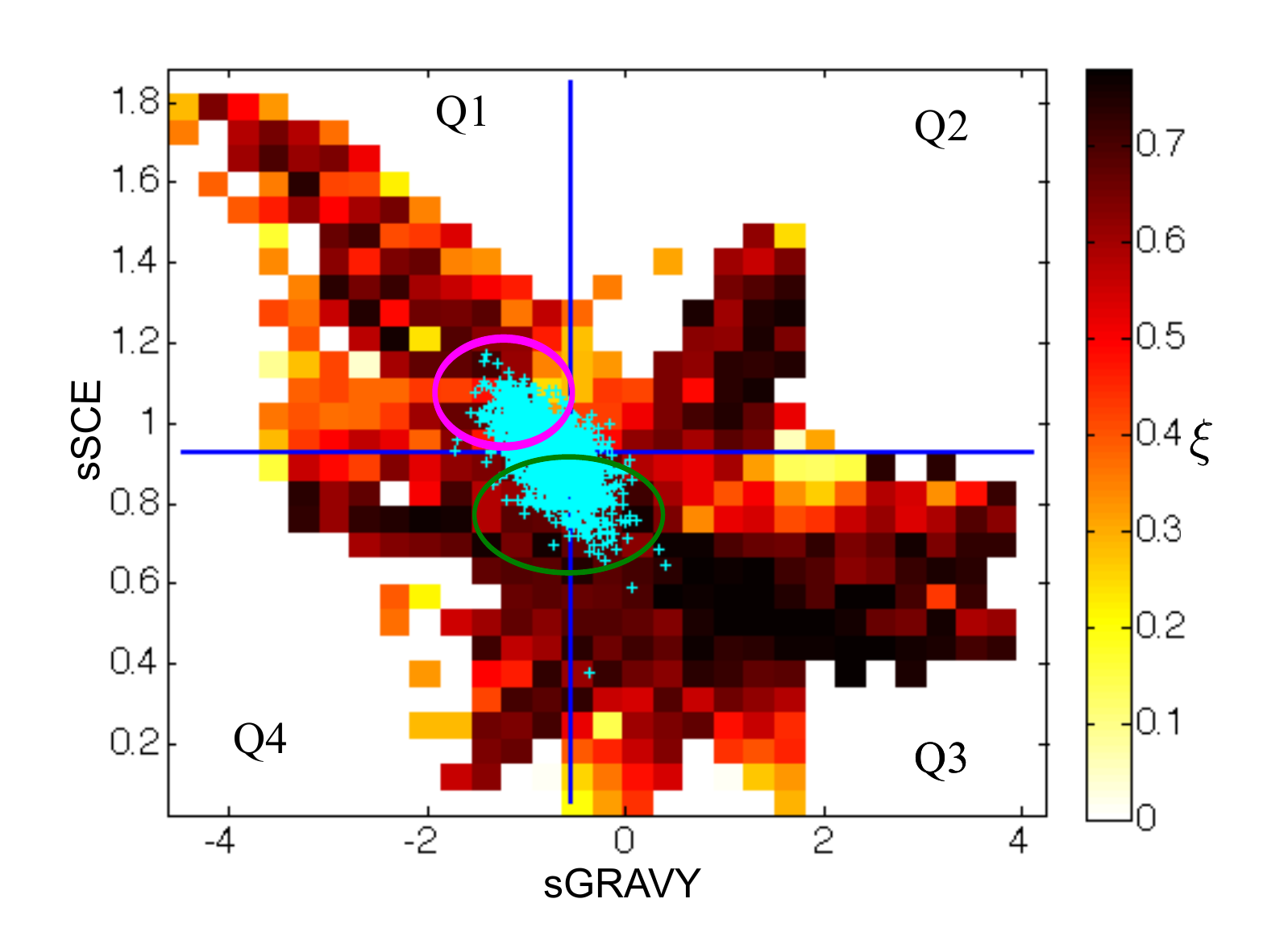}
\caption{Projection of the average predicted propensity $\xi$ over the sGRAVY-sSCE plane. Blue lines divide the plane in four quadrants (Q1, Q2, Q3, and Q4) based on the average sGRAVY and sSCE of the proteins in the \nesg\ database. Cyan pluses represent the projection of the 1,612 structures downloaded from the PDB, whose distribution is used to generate Tables~\ref{tab:easy} and~\ref{tab:crystal_contacts}. The green circle broadly indicates the region where the sSCE is the main driving force to crystallization, while the fuchsia circle indicates the region where specific chemical complementarity plays a more significant role. The model also captures the reduction in propensity associated with limited protein solubility at very low sSCE.}
\label{fig:sce_gravy}
\end{figure}

The complete GPR model also reveals the presence of crystallization hot spots, i.e., regions of $\mathbf{x}$ that give a high $\xi$. In order to locate these hot spots, we select the proteins from the \nesg\ sample in the top 5 percentile for crystallization propensity ($\xi>\xi_{95\%}=0.1$) as starting points for searching the protein property space. We specifically explore how $\xi$ changes when moving away from these starting points by changing the surface residue composition (see Method section for search details). Figure~\ref{fig:sce_gravy} presents the hot spots projected on the sSCE and surface hydropathy (sGRAVY) plane. These two variables are thought to strongly influence the ease with which a protein crystallizes~\cite{janin:1995b,rodier:2005,derewenda:2009}.
We divide the plane in quadrants using the \nesg\ dataset's average sSCE and sGRAVY as delimiters. Interestingly, the model predicts high $\xi$ regions in all four quadrants, but some of these regions are not biologically reasonable for actual proteins and should therefore be discarded. The crystallization propensity $\xi$ is indeed a mathematical object with no physical constraints, and hence can be defined for any combination of properties $\mathbf{x}$. Propensity maxima that are ``unbiological'', such as proteins whose surface is constituted of a single amino acid, can therefore be found.
To obtain an estimate of the property range corresponding to actual proteins, we locate on the sGRAVY-sSCE plane a set of 1,619 distinct monomeric proteins, i.e., proteins whose crystal contacts are not biologically-driven, reported in the PDB that (as of October 2013) were high-resolution ($<2$\AA) and had less than 90\% sequence similarity. 
As shown in Figure~\ref{fig:sce_gravy}, these proteins cover only a small region of the sGRAVY-sSCE plane, locating only two biologically-relevant high-propensity regions. First, a high propensity region spans Q3 and Q4, corresponding to proteins with lower than average sSCE and moderate hydrophobicity (green circle). Second, a series of hot spots are detected in Q1 (fuchsia circle). The landscape roughness of this second region, however, suggests that sSCE and sGRAVY alone do not fully characterize its properties. Additional structural variables need to be considered.
 
In order to get a clearer physico-chemical understanding of the high crystallization propensity regions, we use the GPR model to predict $\xi$ for each of the 1,619 PDB proteins described above, sorted according to the sGRAVY-sSCE quadrant to which they belong (46\% in Q1, 5\% in Q2, 19\% in Q3, and 30\% in Q4). This richer dataset provides a stronger signal than the \nesg\ database alone, but relies on the assumption that these 1,619 proteins do not deviate too strongly from the \nesg\ protein set. In support of this assumption, we note that the distribution of protein properties of the two sets do not significantly differ. From the quadrant analysis, we obtain the distribution of surface residues for proteins that are predicted to be easy ($\xi\geq\tilde{\xi}_{75\%}=0.1$) and hard ($\xi\leq\tilde{\xi}_{25\%}=0.01$) to crystallize within different quadrants, save for Q2, which is too sparsely populated. (Thresholds with a tilde refer to the distribution of modeled crystallization propensity for the set of 1,619 PDB proteins and not to results of the \nesg\ dataset.) Kolmogorov-Smirnov tests determine whether the distributions of protein properties are significantly different (p-value$<0.01$) between easy and hard to crystallize proteins within a given quadrant (Table~\ref{tab:easy}). 
As expected, in Q3 hydrophobicity emerges as the major drive to crystallization. High propensity proteins are also depleted in small residues compared to their recalcitrant counterparts in the same quadrant, as found above. Because these proteins are very hydrophobic, low sSCE results in proteins that are insoluble.
Easy to crystallize proteins in Q1 are enriched for polar residues and for pairs of side chains that involve polar residues. These proteins thus likely rely on electrostatic interactions to form some of their crystal contacts. Surprisingly, they are also depleted in negative residues, which, as discussed above, may be an artifact of the choice of solution conditions. 

   \begin{table}[ht]
\begin{tabular}{|c||c|c|c|c|}
\hline
-&\bf{Q1}&\bf{Q3}&\bf{Q4}\\
\hline
enriched&P,PP,PH&H,HH&\\
\hline
depleted&$-$,S$-$,$-$H,$--$&S,SP,+P&$-$P\\
\hline

\end{tabular}
\caption{ For the different quadrants, the list of properties for which easy to crystallize proteins ($\xi\ge\tilde{\xi}_{75\%}$) are enriched for (or depleted in) compared to hard to crystallize proteins ($\xi\le\tilde{\xi}_{25\%}$). Symbols as in Fig.~\ref{fig:length_scales}.} 
\label{tab:easy}
 \end{table}

Our analysis suggests that two distinct physico-chemical mechanisms drive crystallization, depending roughly on whether a protein has lower or higher than average sSCE. The first mechanism is based on limited interference from side chain entropy for hydrophobic interactions, and the second relies on the formation of specific, complementary charge and polar interactions. To test for the presence of these distinct crystallization mechanisms in an independent way, we analyzed protein crystal contacts (see definition in Methods). These contacts carry the structural signature of the interactions that drive crystal formation. A Kolmogorov-Smirnov test revealed marked differences in residue and residue pair distributions for proteins belonging to different quadrants, especially between the low-sSCE proteins in Q1 and the high-sSCE proteins in Q3 and Q4 (Table~\ref{tab:crystal_contacts}). Q1 proteins are enriched for charged and polar residues, while Q3 and Q4 proteins are enriched for small and hydrophobic residues. Q3 and Q4 can be further distinguished from each other by the higher frequency of hydrophobic (in Q3) vs polar (in Q4) residues present in crystal contacts. Given these results we confirm that Q1 proteins crystallize mostly using complementary electrostatic interactions (enriched for positive-negative pair residues), Q3 using backbone-backbone (enriched for small-small pair residues) and hydrophobic interactions, and Q4 using both backbone-backbone and polar interactions. 
The fact that such a distinction can be  made is particularly remarkable because different crystal contacts of a protein may, in general, involve more than one type of interactions~\cite{fusco:2013b}, which, through averaging, should weaken the statistical signature of this effect.

\begin{table*}[ht]
\begin{tabular}{|c||c|c|c|c|}
\hline
enriched&\bf{Q1}&\bf{Q3}&\bf{Q4}\\
\hline
\bf{Q1}&&+,$-$,++,+$-$,+P,+H,$--$,$-$P,$-$H&+,$-$,++,+$-$,+P,+H,$--$,$-$P,$-$H\\
\hline
\bf{Q3}&S,H,SS,S+,S$-$,SP,SH,PH,HH&&H,SS,HH\\
\hline
\bf{Q4}&S,SS,S+,S$-$,SP,SH,PP&+,$-$,P,++,+$-$,+P,$-$P,PP&\\
\hline

\end{tabular}
\caption{Enrichment for specific properties of the crystal contacts of the proteins belonging to different sSCE-sGRAVY quadrants. For example, position (Q1,Q3) lists the properties for which proteins in Q1 are enriched compared to proteins in Q3. Note that the pairs here indicate interactions between residues on \emph{different} chains rather than neighboring residues on the same chain. Symbols as in Fig.~\ref{fig:length_scales}.}
\label{tab:crystal_contacts}
 \end{table*}

From the analysis above, we note (i) that reducing sSCE is not the only pathway to generate less recalcitrant mutant, and (ii) that proteins with low sSCE are not necessarily easier to crystallize. Our findings suggest that, depending on the level of sSCE, proteins crystallize using two different sets of physico-chemical mechanisms. At high sSCE, crystallization relies mostly on the enthalpic gain of forming favorable electrostatic interactions, such as salt-bridges or polar interactions; at low sSCE, the reduced entropic cost of freezing small residues as well as the hydrophobic effect appear to be the driving forces. 
For the latter group of proteins, if reducing solubility is necessary to form protein crystals, the mutagenic strategy proposed by SER is more likely to be successful. For proteins with higher sSCE, however, too many mutations may be necessary to reach a range of sSCE that promotes crystallization. Mutating a few selected residues that can trigger electrostatic interactions may then be a more effective strategy. For example, replacing small residues with polar residues or mutating charged side chains that are found next to oppositely charged side chains could help promote inter-protein electrostatic interactions.

\subsection{GPC: Solution conditions for different crystallization mechanisms}

In order to test whether the two main crystallization mechanisms identified above are optimized by distinct sets of solution conditions, we trained four separate GPC models on the \nesg\ dataset. Separately for Q1 and Q3, we considered proteins with higher ($\xi>\xi_{50\%}=0.04$) and lower ($\xi<\xi_{50\%}$) than average propensity. 
We were particularly interested in the solution conditions that help crystallize recalcitrant proteins from the Q1 (electrostatic mechanism) and the Q3 (SER/hydrophobic mechanism) quadrants.
The relative abundance of statistically significant solution properties in the Q1 model for recalcitrant proteins indicates that the response of Q1 proteins to changes in solution conditions is more complex than that of Q3 proteins.
 This observation is consistent with chemical interactions in Q1 being more heterogeneous and thus responding to specific solution conditions. In this context, the ionic strength and the presence of high-valency ions (with charge $\pm$2 or $\pm$3) seem to play particularly important roles (Table~\ref{tab:cond_length}). The capacity of certain high-valency ions to coordinate proteins at crystal contacts~\cite{zhang:2008} or to bind proteins active sites may partly explain this sensitivity.

 \begin{table}[ht]
\begin{tabular}{|c|c||c|c|}
\hline
\multicolumn{2}{|c||}{\bf{Q1, $\xi<\xi_{50\%}$}}&\multicolumn{2}{|c|}{\bf{Q3, $\xi<\xi_{50\%}$}}\\
\hline
property&$l$&property&$l$\\
\hline
IS&1.00&\color{blue}{Cadmium}&1.00\\
PEG 10000&1.00&\color{blue}{Nickel}&1.00\\
\color{red}{Citrate}&1.00&\color{blue}{Cesium}&1.00\\
PEG 5000&1.00&\color{red}{Pyrophosphate}&1.09\\
\color{blue}{Strontium}&1.00&\color{red}{Succinate}&1.11\\
\color{blue}{Nickel}&1.03&\color{red}{Iodide}&1.20\\
\color{red}{Tetraborate}&1.04&\color{blue}{Barium}&1.45\\
PEG 200&1.04&\color{red}{Fluoride}&1.47\\
\color{red}{Tartrate}&1.05&\color{blue}{Samarium}&1.73\\
\color{red}{Triphosphate}&1.05&\color{blue}{Copper}&1.84\\
\color{blue}{Samarium}&1.06&PEG 200&1.84\\
\color{blue}{Copper}&1.12&\color{red}{Tetraborate}&1.87\\
\color{blue}{Iron}&1.13&\color{blue}{Iron}&1.99\\
\color{red}{Iodide}&1.19&PEG 2000&2.21\\
\color{blue}{Barium}&1.49&\color{red}{Triphosphate}&2.34\\
\color{blue}{Cesium}&1.64&PEG 5000& 2.64\\
\color{red}{Pyrophosphate}&1.68&\color{blue}{Zinc}&3.13\\
\color{blue}{Cadmium}&1.72&\color{red}{4-aminosalycilate}&4.00\\
PEG 1500&2.49&\color{blue}{Gadolinium}&6.38\\
PEG 550&2.56&\color{red}{Cacodylate}&7.44\\
\hline

\end{tabular}
\caption{Most significant condition properties for hard to crystallize proteins belonging to Q1 and Q3. Properties are colored according to their classification: cation (blue), anion (red), PEG and others (black).}
\label{tab:cond_length}
 \end{table}

\begin{figure}[htb]
\centering
\includegraphics[width=0.5\textwidth]{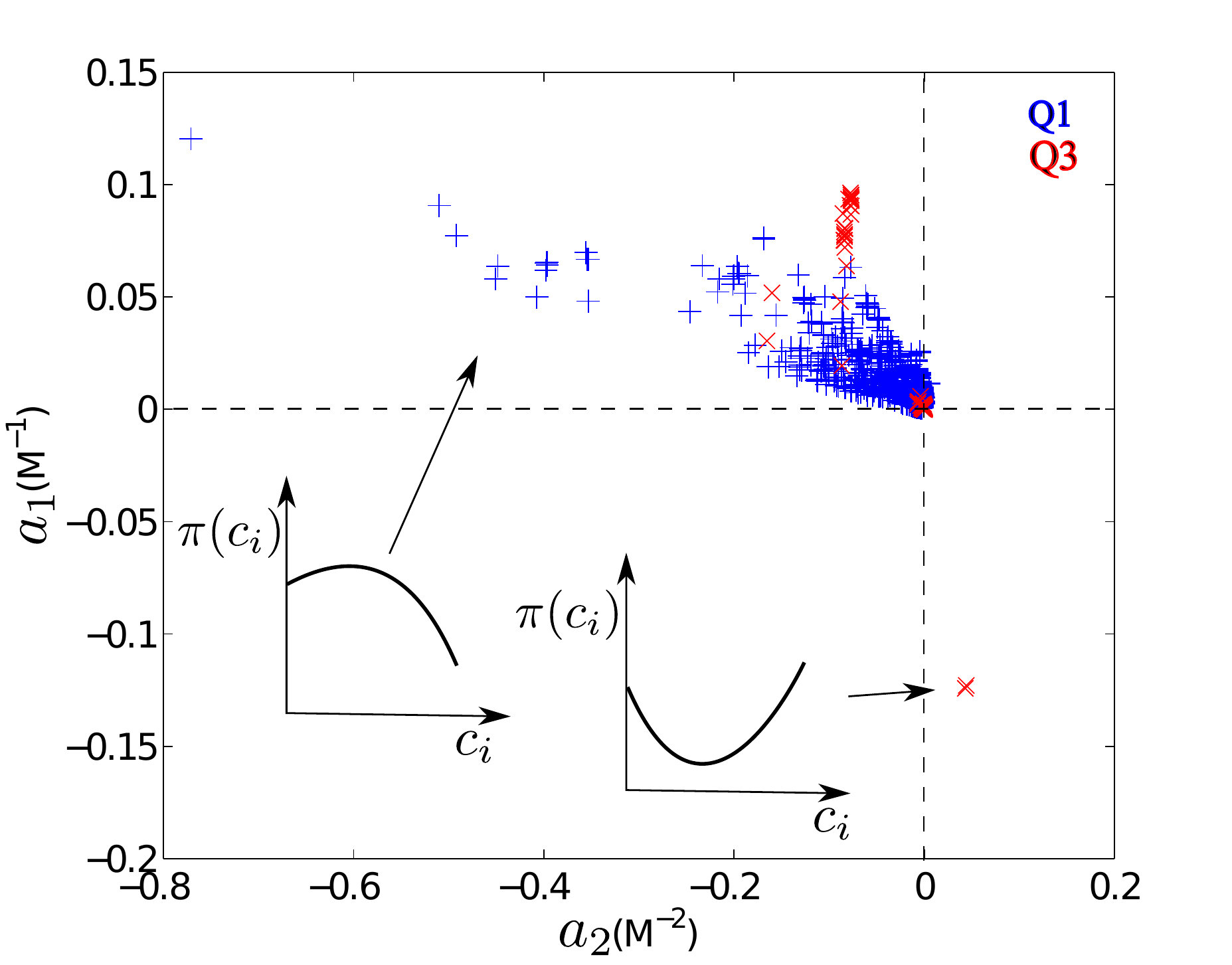}
\caption{Scatter plot of polynomial parameters that characterize the non-monotonic trends in crystallization probability with concentration for different salts. Proteins belonging to Q1 are represented by blue pluses and those belonging to Q3 by red crosses. The insets sketch the probability trend as a function of salt concentration for different combinations of $a_1$ and $a_2$ ($a_2<0$ and $a_1>0$ vs $a_2>0$ and $a_1<0$).}
\label{fig:2poly}
\end{figure}

Even more compelling evidence for the heightened importance of the crystallization conditions in Q1 compared to Q3 is the crystallization response to changes in solute concentration (Fig.~\ref{fig:2poly}). Trained GPC models allowed us to extend the results reported in the \nesg\ database and to explore how the probability of successful crystallization, $\pi$, is affected by the solution features.
Although predictions for solution conditions that are very different from those experimentally tested have a high uncertainty and are essentially meaningless, for conditions similar to those used in the training set, the model may actually enrich that information. We specifically consider the predictions based on the molarity $c_i$ of a given additive $i$ (see Methods).
 As a first approximation, we fit each of these trends to a second-order polynomial $\pi(c_i)=a_0+a_1c_i+a_2c_i^2$. By definition, $a_0$ is the probability that the protein crystallizes without additives, while $a_1$ and $a_2$ qualify the crystallization probability dependence on the additive concentration. The behavior of the model can be visualized as a scatter plot where each symbol represents the response of a given protein to a given salt (Fig.~\ref{fig:2poly}). 
(The lower-left quadrant is empty because monotonically decreasing trends are excluded from this analysis.)
The upper-left quadrant corresponds to proteins whose probability to crystallize is highest at intermediate concentration, and the lower-right quadrant to those that crystallize more easily under either very high or very low additive concentrations (insets in Fig.~\ref{fig:2poly}). (Recall that the underlying data is based on the visual observation of crystals in samples where one or more crystallization conditions out of typically many has proceeded to provide structural data. Not every crystal in every experiment is examined by X-ray (or UV imaging) and it is thus possible that crystals forming in high salt conditions are actually salt crystals.) Additives to Q1 proteins span the upper-left quadrant fairly broadly, which suggests a high heterogeneity in optimal additive concentrations. Most additives to Q3 proteins tend to have small values of $a_2$ and positive $a_1$, corresponding to a nearly linear response. 
Interestingly, many more salt types result in non-decreasing trends for Q1 than for Q3 proteins, i.e., there are more blue than red symbols. Tuning salt type and concentration is thus likely to be a more effective strategy to crystallize proteins in Q1 (electrostatic mechanism) than those in Q3 (SCE/hydrophobic mechanism). In other words, for Q1 proteins tuning crystallization conditions should be sufficient to obtain a crystal, whereas a target protein belonging to Q3 may need more invasive mutagenic approaches if it does not crystallize from standard screens. Reciprocally, if a Q3 protein crystallizes in a standard screen, it is likely to produce hits in several conditions. Interestingly, the two highly-crystallizable \nesg\ proteins, which crystallized in more than 30\% of conditions (PDB ids: 2OYR and 2PGX, Fig.~\ref{fig:protein}), belong to this category.  

\begin{figure*}[htb]
\centering
\includegraphics[width=\textwidth]{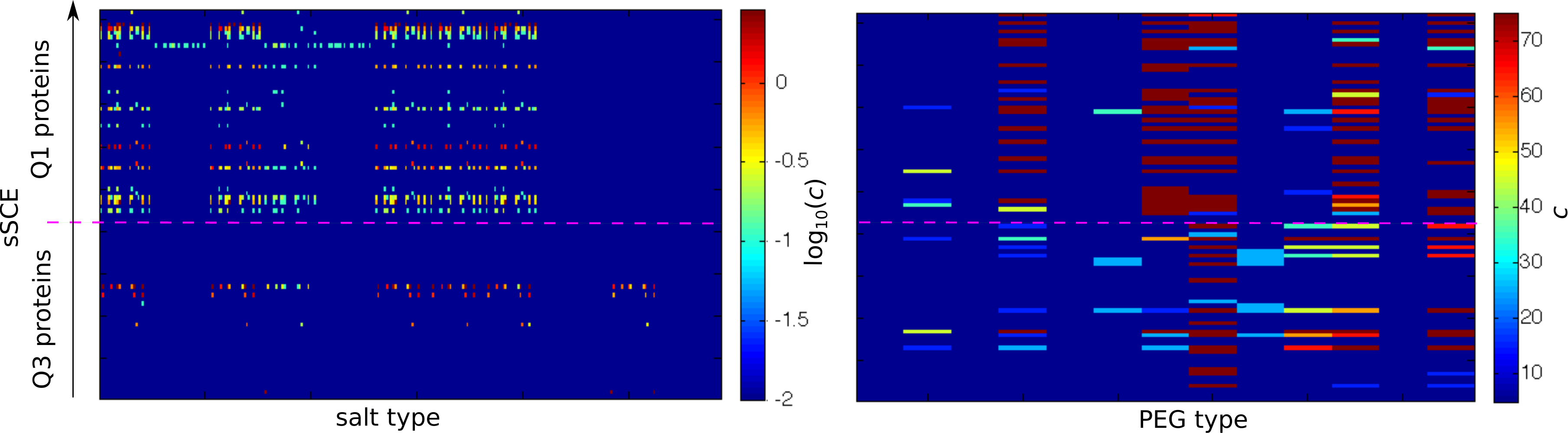}
\caption{Optimal conditions for low crystallization propensity proteins in Q1 (upper side) and Q3 (lower side) for various salt types (molar concentration) and PEG types (\% mass concentration). Proteins are ordered by sSCE; salt types are ordered by cation (ammonium, calcium, lithium, magnesium, manganese, potassium, rubidium, sodium, zinc, barium, cesium, cobalt, copper, iron, gadolinium, nickel, samarium, strontium, cadmium) 
and, within each cation, by anion (citrate, malonate, succinate, tartrate, acetate, bromide, cacodylate, carbonate, chloride, citrate tribasic, fluoride, formate, iodide, molybdate, nitrate, phosphate monobasic, phosphate dibasic, phosphate tribasic, pyrophosphate tetrabasic, sulfate, tetraborate, thiocyanate, thiosulfate, 4-aminosalicylate); and PEG types are ordered by molecular weight (200, 400, 550, 1000, 1500, 2000, 3350, 4000, 5000, 6000, 8000, 10000, 20000 g/mol).}
\label{fig:conditions}
\end{figure*}

The model's optimal salt and PEG concentration for each hard to crystallize protein in Q1 and Q3 varies by quadrant (Fig.~\ref{fig:conditions}). For Q3 proteins, adding salt rarely improves the probability to crystallize (4 out of 42 proteins) and, when it does, high salt concentrations are preferred.
By contrast, the crystallization probability of more than 40\% of Q1 proteins is improved by the presence of salt (at concentrations between 0.1 and 1 M). Once again, these predictions support an electrostatics-dominated mechanism for Q1. Most of the optimal conditions for these proteins cluster by salt type (contiguous patterns along the horizontal axis). Because salt types are ordered by cation, the clustering of the results suggests that the crystallization of Q1 proteins is more sensitive to the type of cation than to the type of anion. PEG also results in distinct crystallization patterns for Q1 and Q3 proteins. The former prefer high concentrations of large PEG molecules, while the latter heterogeneously respond to the presence of PEG, both size- and concentration-wise. 
These results suggest that the successful crystallization of Q1 proteins requires a wide sampling of salt types (specifically cations) and concentrations. For these proteins, it may thus suffice to tune the crystallization conditions without resorting to mutagenesis.
By contrast, tuning the type and concentration of PEG appears to be more effective for Q3 proteins, which are, however, generally less sensitive to solution conditions. Mutations, such as those suggested by SER, may then be necessary to promote crystallization.

\section{Conclusion}
Using state-of-the-art statistical techniques on a detailed database of protein crystallization experiments coupled with extensive information on those proteins and their resulting structures, our study recapitulates, \emph{with a single model}, many physico-chemical factors that independent studies have related to crystallization propensity, and detects the correlations between these variables. In addition, our model distinguishes two main mechanisms that drive monomeric protein crystal assembly. One is mainly entropic and exploits low side chain entropy and hydrophobicity; the other is energetic and relies on complementary electrostatic interactions.  The key contribution from electrostatic interactions provides further evidence that crystal contacts 
have a specific physico-chemical signature even if they are not biologically functional~\cite{janin:1995,janin:1995b,carugo:1997,zhuang:2011,wilkinson:2004,fusco:2013b}. These interactions are indeed of the same nature as those that traditionally result in specific and thus biologically relevant interactions, such as protein complex assembly or protein-target recognition. The knowledge accrued over the years for these interactions~\cite{Jones:1996} may thus be useful for understanding and designing crystal contacts~\cite{Banatao:2006,Lanci2012}.

The GP-based models developed in this study also estimate the crystallization propensity of any protein, given a set  of its physico-chemical properties, and identify mutagenesis strategies that are more likely to yield protein crystals. For example, we find that it may be favorable to mutate positive-negative surface residue pairs to uncharged residues or small residues to polar ones, in order to crystallize a recalcitrant Q1 protein, whereas SER guidelines may be more useful for crystallizing Q3 proteins. In addition, using data from crystallization screens, an improved set of solution conditions can be determined given some of the protein surface properties. For example, fine-tuning salt concentration and cation type appears to be an effective strategy for proteins with higher than average sSCE. In contrast, using a high salt concentration and the addition of PEG appear to be more effective approaches for crystallizing proteins with lower than average sSCE. 

Although our analysis cannot be directly applied to \textit{de novo} protein crystallization, a coarse Q1/Q3 classification may still be possible based on a protein's average SCE, which linearly correlates to its sSCE and can be determined from the primary structure. This approximate assignment may narrow down which one of the two main crystallization approaches is more likely to be successful.
More precise and complete structural information, e.g., residue types and pairings, could also be obtained by combining different (imperfect) protein folding algorithms~\cite{Dill:2012}. For example, relatively precise estimates of sSCE can be calculated from available computational tools, such as PredictProtein~\cite{rost:2004}. It should thus be possible to compute from sequence information alone what residues are likely to be exposed and, consequently, to estimate the protein properties that the GPR and the GPC models need to predict its crystallization propensity and optimal crystallization conditions.
Future studies will integrate the current models with algorithms that estimate these properties, and assess their experimental success. 

Finally, the accuracy of any statistical model depends on the quantity and quality of the training set. Our findings emphasize the need for an increased availability and standardization of protein crystallization datasets~\cite{newman:2012}. A richer characterization of the experimental outcomes would also extend the reliability of these models. For example, different successful crystallization conditions can yield distinct crystal forms and thus crystal contacts for the same protein. 
The availability of crystal symmetry and contact information for different conditions would refine our understanding of the correlation between  experimental conditions and the (solution mediated) protein-protein interactions that drive crystallization. 
Similarly, unsuccessful conditions could be defined more finely depending on whether a protein remained soluble or gelled. Interpreting this data in light of phase diagrams would further clarify the physico-chemical basis for protein crystallization and guide future experiments. It is thus reasonable to anticipate that the extension of statistical models and the increased availability of training datasets will help guide biomolecular crystallization toward a more rational basis.

%
\section{Methods}\label{sec:method}

\subsection{Data}\label{sec:data}

The crystallization database reports binary crystallization outcomes in 198 samples of 182 unique proteins from the \nesg\ (list of PDB IDs in Supplementary Materials) each in 1,536 solution conditions in microbatch under-oil experiments conducted at the Hauptman-Woodward Medical Research Institute High-Throughput Crystallization Screening (HTS) laboratory. The concentration of the various chemicals, proteins, and pH are reported. The solution conditions span six generations (generations 5 to 9) of the cocktails used in the HTS center with approximately half the conditions representing commercially available crystallization screens and the other half an incomplete factorial sampling of chemical space~\cite{luft:2011}. Most experimental conditions fall into two categories: moderate to high salt alone, and low salt with PEG representing typical crystallization strategies. Although a total of 311 different chemicals are used, we focused on the effect of ions (divided in 19 cations and 24 anions) and 13 types of PEG for a total of 56 analyzed chemical species. The pH distribution is slightly biased towards lower values (mean pH $=6.8$). The chemical species concentrations are combined to obtain the ionic strength of the solution (IS), a Hofmeister series coefficient ($\mathrm{HS}_a$ for anions and $\mathrm{HS}_c$ for cations) and a depletion effect coefficient (DEP),
\begin{eqnarray}
\mathrm{IS}&=&\frac{1}{2}\sum_{i\in \mathrm{ions}}c_iZ_i^2\\
\mathrm{HS}_c&=&\sum_{i\in\mathrm{cations}}c_i \mathrm{hs}_i\\
\mathrm{HS}_a&=&\sum_{i\in\mathrm{anions}}c_i \mathrm{hs}_i\\
\mathrm{DEP}&=&\sum_{i\in\mathrm{PEG}}\frac{c_i}{R_i^3}(R_p+R_i)^3\nonumber\\
&=&\sum_{i\in\mathrm{PEG}}\frac{c_i}{M_i^{3/2}}(R_p+M_i^{1/2})^3,
\end{eqnarray}
where $c_i$ is the species concentration, $Z_i$ the ion charge, $R_p$ is the solvated protein radius of gyration, $M_i$ is the PEG molecular mass~\cite{oosawa:1954,Vrij:1976,Dijkstra:1998,lee:2009}, and $\mathrm{hs}_i$ is a Hofmeister index that ranks the species from more to less kosmotropic (cations: ammonium, cesium, rubidium, potassium, lithium, sodium, barium, magnesium, manganese, zinc, cadmium, calcium, cobalt, copper, nickel, strontium, iron, gadolinium, samarium; anions: triphosphate, tricitrate, sulfate, tartrate, carbonate, thiosulfate, diphosphate, succinate, citrate, acetate, malonate, fluoride, formate, chloride, bromide, iodide, monophosphate, thiocyanate)~\cite{zhang:2006}.

It is important to note that the data comes from samples that produced hits in crystallization screening \textit{and} then went on to yield a structure deposited in the PDB. Crystal hits that yielded no structural data are beyond the scope of our analysis. For the 182 proteins studied, 29\% gave hits in ten or fewer of the 1,536 different chemical cocktails but 18\% gave hits in 100 or more cocktails. Typically only the best set of initial conditions go forward to optimization, hence we have no data on how well a crystal may have diffracted when grown in one of the other solutions. In this analysis we also give equal weight to each crystallization hit, which introduces additional noise in the data. It is also important to note that the protein samples were all prepared in a common buffer, which reduces the number of solution variables. 

PyMol was used to determine the structural characteristics of each protein from its PDB structure: the fraction of the protein surface carrying each residue (a residue was considered exposed if at least 2.5 $\mathrm{\AA }^2$ of its surface is exposed), the solvent accessible surface area (SASA), the radius of gyration, and the isoelectric point (pI). Global and surface values for the grand average of hydropathicity index (GRAVY) (measure of hydropathy)~\cite{kyte:1982}, the polarity (POL) coefficients~\cite{derewenda:2009}, and the side chain entropy (SCE)~\cite{derewenda:2009,doig:1995} were obtained by averaging the value for each residue, respectively in the protein and on the protein surface (Fig.~\ref{fig:protein}). Note that we defined the magnitude of sSCE such that more flexible residues have a higher sSCE, which is opposite to the definition of Ref.~\cite{doig:1995}.

The residues were clustered in categories: small (G, A), positively charged (H, R, K), negatively charged (D, E), polar (C, S, T, N, Q) and hydrophobic (L, I, V, F, Y, M, W, P). To incorporate the first many-body correction, we also determined the number of neighboring (within 5 \AA\ of each other) residue categories (small-small, small-polar, and so on) normalized over the total number of neighboring pairs. These variables were used both in absolute number and weighted by their solvent accessible area, because more exposed residues may play a larger role than less exposed ones in protein crystallization. Note that these predictive variables are not all independent and some have to satisfy certain constraints. In particular, the surface fraction covered by each amino acid type has to sum up to 1, and, given the surface amino acids, sGRAVY, sSCE, and sPOL are uniquely determined.

Combining this information generates two sets of data. The first associates a crystallization propensity (fraction of successful experiments) to each protein characterized by 89 protein features (Fig.~\ref{fig:protein}).
The second reports the success or failure of each experiment for each protein (254,623 experiments in total) characterized by the solution conditions (61 cocktail features) and the protein features for a total of 150 predictive variables. The data is available upon request.

\subsection{Crystal contacts analysis}

Similarly to previous studies~\cite{derewenda:2009}, we defined crystal contacts as the regions on the proteins surface that are within 5 \AA\ from surface residues on a neighboring chain in the protein crystal. To identify the crystal contacts, we used PyCogent~\cite{pycogent}, whose structural biology tool-kit is an extension of PDBZen~\cite{derewenda:2009}. In-house Python scripts classified the properties of each crystal contact.

\subsection{Statistical model}

In standard linear and generalized linear models, the response variable $y$, whether continuous or discrete, is a function $\sigma$ of a linear combination of the predictive variables $\mathbf{x}$
\begin{equation}
y=\sigma(\mathbf{x}^T\mathbf{w})+\epsilon,\label{eq:linear}
\end{equation}
where $\mathbf{w}$ indicates the weights of the variables and $\epsilon$ is the uncertainty of the model. A non-linear dependence among the predictive variables cannot be captured by this framework. Gaussian processes discard the assumption of linearity and place a prior on any possible functional form, giving more flexibility to the model. In contrast to deterministic methods (such as Support Vector Machine), GP are Bayesian, which means that they assign a probability distribution to the response variable and provide a confidence interval on the predicted value. In the following, we briefly summarize GP regression and classification. More details can be found in Ref.~\cite{GP}.

In the simplest version of GP inference, the latent function $f(\mathbf{x})$ replaces the linear dependency in Eq.~(\ref{eq:linear}). The prior on $f$ is
\begin{equation}
p(f|\mathbf{x})\sim N(0,K),
\end{equation}
where $N(0,K)$ indicates a multinormal distribution with zero mean and covariance matrix $K$. Among the several available options for $K$, we opt for the widely used squared exponential, so that
\begin{equation}
K(\mathbf{x}_i,\mathbf{x}_j)= \mathrm{exp}(-\gamma(\mathbf{x}_i-\mathbf{x}_j)P(\mathbf{x}_i-\mathbf{x}_j)^T),
\end{equation}
where $\gamma$ and the diagonal matrix $P$ are (hyper-)parameters that have to be optimized. In particular, each element $p_i$ of the diagonal of $P$ can be related to the typical length scale $l_i$ of variable $i$ as $p_i=l_i^{-1/2}$. Large $l_i$ correspond to less important variables, while a small $l_i$ identifies a variable whose variation strongly affects the response variable. For the scope of this study, we arbitrarily defined variables to be significant if $l<100$, which is roughly the half point between the largest and the smallest length measured in logarithmic scale. 

\subsubsection{GPR}

In GPR, the response variable is defined as $f=f(\mathbf{x})$.  The predictive probability over a test set $\mathbf{x}_{\mathrm{test}}$, given a training set $(\mathbf{x}_{\mathrm{train}},f_{\mathrm{train}})$, is $p(f_{\mathrm{test}}|\mathbf{x}_{\mathrm{test}},\mathbf{x}_{\mathrm{train}},f_{\mathrm{train}})\sim N(\hat{f},\hat{K})$, where
\begin{eqnarray}
\hat{f}&=&m+K(\mathbf{x}_{\mathrm{test}},\mathbf{x}_{\mathrm{train}})K(\mathbf{x}_{\mathrm{train}},\mathbf{x}_{\mathrm{train}})(f_{\mathrm{train}}-m)\nonumber\\
\hat{K}&=&K(\mathbf{x}_{\mathrm{test}},\mathbf{x}_{\mathrm{test}})-\nonumber\\
&&K(\mathbf{x}_{\mathrm{test}},\mathbf{x}_{\mathrm{train}})K(\mathbf{x}_{\mathrm{train}},\mathbf{x}_{\mathrm{train}})K(\mathbf{x}_{\mathrm{train}},\mathbf{x}_{\mathrm{test}}),\nonumber
\end{eqnarray}
in which $m$ is the mean of the observed response variable over the training set. Sampling the predictive distribution provides predictions on the response variable $f$.

The hyper-parameter selection is performed by optimizing the marginal log-likelihood
\begin{eqnarray}
\lefteqn{\mathrm{log}[p(f_{\mathrm{train}}|\mathbf{x}_{\mathrm{train}},\gamma,P)]=}\nonumber\\
&&-\frac{1}{2}[f_{\mathrm{train}}^TK^{-1}f_{\mathrm{train}}-\mathrm{log}|K|-n\mathrm{log}2\pi],\label{eq:marginal_likelihood}
\end{eqnarray}
where $n$ is the sample size. By determining the gradient of the marginal log-likelihood, any conjugate gradient optimization method can be used to locate local maxima.
The marginal log-likelihood was maximized by sampling $p_i$ according to  a beta distribution with a shape parameter of 0.5. For each initial condition of the hyper-parameters, we performed a conjugate gradient search to identify the corresponding local maximum (Fig.~\ref{fig:LOOCV}A).

We used GPR to study the protein crystallization propensity. Because the propensity ranges between 0 and 1 by definition, a brute-force regression is not appropriate ($f$'s domain is the whole real line). A possible solution to the problem is to link crystallization propensity $\xi$ and $f$ using a sigmoidal function. The drawback of this approach is that very low propensity values, which are by definition affected by large relative uncertainty, correspond to large negative values of $f$. As a result, the inference process gives poorer predictions for proteins that are easy to crystallize. Because these proteins are of greatest interest to us, we opted instead for $f=\frac{\xi}{1-\xi}$. Although small nonphysical negative values of propensity are then allowed, this transformation is close to linear for small values of $f$ and emphasizes the contribution of high-propensity proteins. In order to have all the predictive features on a similar scale (each of them spans very different ranges of values), we scaled each feature according to their mean and standard deviation (z-scores). Length scales $l_i$ then correspond to the actual significance levels of each property.

The search for hot spots, which identifies features that maximize protein propensity, was performed by laying down a grid over the feature space with a fineness that depended on the length scale of the corresponding dimension. For variable with $l_i<100$, four equidistant points in the physical range were used, and otherwise only the mid value was used. Although not exhaustive, trials with finer grids did not detect additional maxima.

\subsubsection{GPC}

In GP binary (success/failure) classification, the probability of success $\pi$ is connected to the latent function $f$ by
\begin{equation}
\pi(\mathbf{x})=\phi[f(\mathbf{x})],
\end{equation}
where $\phi$ is a sigmoidal function, such as logistic or probit. A prediction for $\pi$ can be obtained in two steps. First, the distribution of the latent variable $f$ over a test case has to be computed using a training set $(\mathbf{x}_{\mathrm{train}},y_{\mathrm{train}})$, where $\mathbf{x}_{\mathrm{train}}$ indicates the explanatory variables values in the set and $y_{\mathrm{train}}$ the corresponding success/failure outcome. 
\begin{eqnarray}
\lefteqn{p(f_{\mathrm{test}}|\mathbf{x}_{\mathrm{train}},\mathbf{x}_{\mathrm{test}},y_{\mathrm{train}})=}\nonumber\\
&&\int p(f_{\mathrm{test}}|\mathbf{x}_{\mathrm{test}},f)p(f|\mathbf{x}_{\mathrm{train}},y_{\mathrm{train}})df,\nonumber
\end{eqnarray}
where the posterior distribution is $p(f|\mathbf{x},y)=p(y|f)p(f|\mathbf{x})/p(y|\mathbf{x})$. Second, the probabilistic prediction is obtained
\begin{eqnarray}
\lefteqn{\pi(\mathbf{x}_{\mathrm{train}},\mathbf{x}_{\mathrm{test}},y_{\mathrm{train}})=}\nonumber\\
& & \int \phi(f_{\mathrm{test}}) p(f_{\mathrm{test}}|\mathbf{x}_{\mathrm{train}},\mathbf{x}_{\mathrm{test}},y_{\mathrm{train}}) df_{\mathrm{test}}.\nonumber
\end{eqnarray}

Unlike for GPR, these integrals cannot be simplified because of the non-Gaussian form of $\phi$. As a result, either analytical approximations or numerical methods must be used. In this study, the problem is further complicated by the large size of the dataset (each sample corresponds to a different experiment), which makes any computation involving the GP prior matrix $P$ intractable. To bypass this problem, we adopted the sparse approximation method implemented in Ref.~\cite{IVM} (Informative Vector Machine), which relies on incremental Gaussian approximations of the posterior distribution to provide parameter optimization for a probit GP classification.

To determine the best classification model, we constrained the parameters of the protein properties to their GPR values, and maximized the marginal log-likelihood over the parameters corresponding to the solution conditions. The log-likelihood  maximum search used a conjugate gradient algorithm starting from different initial values sampled according to a beta distribution with shape parameter 0.5.

In the GPC analysis reported in the Results section, we focused on how $\pi$ is affected by the concentration of each individual additive. In this case, for a given additive $i$ of concentration $c_i$, we determined $\pi(c_i)=\pi(\mathbf{x}_i)$, where the solution feature vector $\mathbf{x}_i$ corresponds to a condition with neutral pH, additive $i$ concentration set to $c_i$, and all the other additive concentrations set to zero. The ionic strength (IS), Hofmeister series parameters (HS$_c$ and HS$_a$), and the depletion parameter (DEP) were then determined given additive $i$'s properties and concentration as defined in Equations (1), (2), (3), and (4).

\section*{Acknowledgments}
We thank Prof. Gaetano Montelione from \nesg\ for supplying samples for crystallization screening at the Hauptman-Woodward medical Research Institute. Sample preparation and data acquisition were supported in part by the Protein Structure Initiative of the National Institutes of Health, NIGMS grant U54 GM094597. Crystallization screening results and meta data was supplied from the University of Buffalo Center for Computational Resources (CCR) through research supported by NIH R01GM088396 (EHS, JRL, and AEB). DF and PC acknowledge support from National Science Foundation Grant No. NSF DMR-1055586. TJB acknowledges REU support from National Science Foundation Grant NSF CHE-1062607.

\bibliography{buffalo}

\bibliographystyle{prsty} 

\end{document}